\documentclass[11pt]{article}
\usepackage{ourstyle}
\usepackage{authblk}

\title{Threshold-Based Graph Reconstruction Using Discrete Morse
  Theory\footnotemark[1]}

\author[1,2]{Brittany Terese Fasy}
\author[3]{Sushovan Majhi}
\author[4]{Carola Wenk}
\affil[1]{School of Computing, Montana State University}
\affil[2]{Department of Mathematical Sciences, Montana State University}
\affil[3]{Department of Mathematics, Tulane University}
\affil[4]{Department of Computer Science, Tulane University} \date{}

\begin{document}

\twocolumn[{
    \maketitle
    \vspace*{-7ex}
    \begin{addmargin}{7em}{7em}
      \begin{center}
        A\scshape{bstract}
      \end{center}
      Discrete Morse theory has recently been applied in metric graph
      reconstruction from a given density function concentrated around an
      (unknown) underlying embedded graph.  We propose a new noise model for the
      density function to reconstruct a connected graph both topologically and
      geometrically.
    \end{addmargin}
    \vspace*{1ex}
}]

\section{Introduction}

\footnotetext[1]{The authors would like to acknowledge the generous support of
  the National Science Foundation under grants CCF-1618469 and CCF-1618605.}

Graph-like or filamentary structures are very common in science and
engineering. Examples include road networks, sensor networks, and earthquake
trails.  With the advent of modern sampling techniques, very large amounts of
data, sampled around such (often hidden) structures, are becoming widely
available to data analysts.

\paragraph{Problem Statement.}
Given a set of points sampled around an unknown metric graph~$G$
embedded in $\R^2$, output a metric graph $\hat G$ that has the same
homotopy type as $G$ and has a small Hausdorff distance to~$G$.

\paragraph{Background.}
Graph reconstruction from noisy samples has been studied extensively in the last
decade; see e.g.,
\cite{dey_graph_2018_socg,ge_data_11,wang_efficient_2015,Ahmed:2015:CTD:2820783.2820810}.
One can typically classify noise models for reconstruction problems into two
categories: Hausdorff noise and non-Hausdorff noise.  A sample $S$ may not lie
exactly on $G$, however $S$ is sampled from a very small offset of $G$. In this
case, the Hausdorff distance between the sample and ground truth is assumed to
be very small. We call such a noise \emph{Hausdorff} noise. The situation
becomes different in the presence of outliers in $S$. If outliers in $S$ are far
away from $G$, they contribute to an uncontrollably large Hausdorff distance. In
this paper, we aim at geometric reconstruction of Euclidean graphs under
\emph{non-Hausdorff} noise model.

Several non-Hausdorff based graph reconstruction approaches use the density of
the sample points in the ambient space. In this density-based reconstruction
regime, a density function over a rectangular grid of pixels in the plane is
computed from the raw sample $S$. There are several ways one can define density
on the planar grid. A histogram computation or a kernel density estimate are
usually very popular and easy to implement in practice. Then, an appropriate
threshold is chosen to get a thickened graph as the super-level set of the
density at the threshold. Some algorithms work by choosing this threshold
empirically, whereas some others, e.g., \cite{Ahmed:2015:CTD:2820783.2820810},
use systematic topological techniques like persistent homology to choose a set
of thresholds just big enough to capture the desired topological changes in the
sub-level set filtration dictated by the density function. While most of the
previous approaches gained success in practice, not much has been proved
theoretically to guarantee the desired topological or geometric
correctness. Also, the output is usually a region around the underlying
graph. Finally, one prunes the region to extract a graph like structure from it
using some heuristic thinning algorithm.

\paragraph{Related Work. }
Our work is inspired by the recent work by Dey et
al.~\cite{dey_graph_2018_socg}. The authors use a topological technique called
discrete Morse theory to extract the cycles of the underlying graph~$G$ from a
density function. They show that if the density function satisfies a noise
model, that the authors call an $(\omega, \beta,\nu)$-approximation, then the
output $\hat G$ of their algorithm has the same homotopy type as $G$. However,
the noise model is too simplistic to capture degree one vertices of $G$. For
this reason, the leaves or the ``hairs'' of $G$ cannot be reconstructed,
resulting in a large (undirected) Hausdorff distance between $\hat G$ and $G$.

\paragraph{Our Contribution. }
In order to overcome the above mentioned limitations of the algorithm developed
in \cite{dey_graph_2018_socg}, we propose a two-threshold based noise model for
the density function that is more practical and that can localize all vertices
of $G$. Using different thresholds for the graph vertices and the graph edges,
we develop an algorithm (\algref{alg:main}) that can output a reconstruction
that is also geometrically close to $G$. We prove in \thmref{thm:hom} that the
output of our algorithm successfully captures both the topology and geometry of
the underlying graph $G$.

\section{Discrete Morse Theory}
Let $K$ be a finite simplicial complex. A \emph{discrete vector field} $V$ on
$K$ is a collection of pairs $(\tau^{(p)},\sigma^{(p+1)})$ of simplices of $K$
such that $\tau^{(p)}<\sigma^{(p+1)}$ and each simplex of $K$ appears in at most
one of such pair. Here, the symbol~`$<$' denotes the face relation and the
superscript denotes the dimension of the simplex. A simplex $\sigma\in K$ is
called \emph{critical} if $\sigma$ does not take part in any pair.  We define a
V-path as a sequence of simplices
$$\tau_0^{(p)}, \sigma_0^{(p+1)}, \tau_1^{(p)}, \sigma_1^{(p+1)}, \ldots,
\sigma_r^{(p+1)}, \tau_{r+1}^{(p)},$$ where $r>0$,
$(\tau_i^{(p)},\sigma_i^{(p+1)})\in V$ and $\tau_{i+1}^{(p)}<\sigma_i^{(p+1)}$
for all $i\in\{0,\ldots,r\}$.  The \emph{Morse cancellation} of a pair of
critical simplices $\tau,\sigma$ takes place when there is a unique V-path from
a co-dimension one face of $\tau$ to $\sigma$. This process of cancellation
reverses the vectors along that V-path to obtain another vector field on
$K$. For more details see \cite{forman_users_02}. Finally, for a critical
simplex $\sigma$ we define its \emph{stable manifold} to be the union of the
V-paths that end at $\sigma$. Similarly, we define its \emph{unstable manifold}
to be the union of the V-paths that start at $\sigma$. For definitions and more
details see \cite{forman_users_02,dey_graph_2018_socg}.

\section{Double Threshold}
For ease of presentation, we define the noise model in the smooth set-up. Let
$\Omega$ be a planar rectangle, and let $G$ be a finite planar graph embedded
inside~$\Omega$. Let $\omega$ be a small positive number such that $G^\omega$,
the $\omega$-offset of $G$, is contained in $\Omega$ and has a deformation
retraction onto $G$. Also, for each vertex~$v$ of $G$, we call the $\omega$-ball
centered at $v$ the \emph{vertex region} of~$v$. Now, let $\mathcal{V}^\omega$
be the union of all vertex regions of~$G$. We call a density function $f$ on
$\Omega$ an $(\omega,\beta_1, \beta_2,\nu)$-approximation of $G$ if
\[
f(x)\in
\begin{cases}
  [\beta_1, \beta_1+\nu],& \text{ if } x\in \mathcal{V}^\omega \\
  [\beta_2, \beta_2+\nu],& \text{ if } x\in G^\omega-\mathcal{V}^\omega \\
  [0,\nu],& \text{ otherwise }
\end{cases}
\]
where $\beta_1>\beta_2+2\nu,\beta_2>2\nu$. In this case, we call $\beta_1$
and~$\beta_2$ the thresholds for $f$. Throughout this paper, we assume our
density function is an $(\omega,\beta_1, \beta_2,\nu)$-approximation. In
practice, these four parameters are unknown. However, in our algorithm, we use a
cut-off $\delta$ such that
\\ $\nu<\delta<\min{(\beta_2-\nu,\beta_1-\beta_2-\nu)}$ and which is assumed to
be known to us.

In order to reconstruct $G$, the density is expected to assume very large values
inside $G^\omega$ relative to the outside region. Here, a small noise or
perturbation $\nu$ has been assumed. The above mentioned two thresholds make
this noise model close to real-world applications involving the extraction of
road-networks from GPS trajectory data. Because points along trajectories make
the density higher near the intersections than than the edges, this noise model
enables us to correctly reconstruct not only the topology but also the geometry
of $G$ as shown in \thmref{thm:hom}.

\section{Algorithm}
We devise our reconstruction algorithm, \algref{alg:main}, by using discrete
Morse cancellation guided by persistence pairs.

\begin{algorithm}[t]\label{alg:main}
  \KwData{The discretized domain $K$, the density function $f$, the threshold
    $\delta$.}

  \KwResult{The reconstructed graph $\hat G$.}

  Initialize $V$ as the trivial vector field on $K$.

  Initialize $\hat{G}=\emptyset$

  Run persistence on the super-level set filtration of $f$ to get the
  persistence pairs $P(K)$.

  \For{$(\sigma,\tau)\in P(K)$ with $Pers(\sigma,\tau)<\delta$} { Try to perform
    a Morse cancellation for the pair.

    Update V.
  }

  \For{$(v,e)\in P(K)$ and $(e,t)\in P(K)$ with persistence $\geq\delta$} {
    $\hat G$ = $\hat G\cup\{\text{ stable manifold of } e\}$.
  }

  output $\hat G$
  \caption{Graph Reconstruction Algorithm}
\end{algorithm}

\paragraph{Analysis of Algorithm}
We start with a discretization $K$ of the planar rectangle $\Omega$. For
example, $K$ can be a planaer two-dimensional cubical complex. Let the density
function $f:K\to\R$ be an $(\omega, \beta_1, \beta_2,\nu)$-approximation and let
cutoff $\nu<\delta<\min(\beta_2-\nu,$$\beta_1-\beta_2-\nu)$. Our goal is to
construct a discrete vector field $V$ on $K$ that is associated to a discrete
Morse function that is much simpler than~$f$. This way, we clean the density
function from the noise administered by $\nu$. We initialize $V$ with the
initial vector field $K$ in which all simplices are critical. In order to remove
non-genuine critical simplices, we run persistence on the super-level set
filtration of $K$ defined by $f$. Then, for each persistence pair
$(\sigma,\tau)$ with persistence smaller than $\delta$, we try to perform Morse
cancellation of the Morse pair $(\sigma,\tau)$ to update $V$. After the
cancellations are done, we get $V$ which is a cleaner discrete gradient field on
$K$. We can show that the resulting $V$ only contains genuine critical points,
i.e., for each graph vertex we have a critical vertex $v$ of $K$ in its vertex
region and for each edge $e$ of $G$ we have a critical edge in $V$. All these
critical vertices and edges will be contained in $G^\omega$. Moreover, these
critical vertices and edges are characterized by their persistence being larger
than $\delta$. Therefore, to extract the edges of $G$ we consider each edge of
$K$ with persistence $>\delta$ and compute their stable manifolds. The union of
their stable manifolds is the reconstruction $\hat G$.

\section{Reconstruction Guarantees}
The two thresholds help us to localize the critical vertices of the discrete
gradient field inside the vertex regions. The output $\hat G$ has the same
homotopy type as $G$ as shown in the following theorem.

\begin{theorem}[Graph Reconstruction]\label{thm:hom}
  If $G$ is a connected, embedded planar graph in a cubical complex $K$ and $f$
  is an $(\omega,\beta_1,\beta_2,\nu)$-\\approximation of $G$ then the output
  $\hat G$ of \algref{alg:main} has the same homotopy type as $G$. Moreover,
  $d_H(G,\hat G)<\omega$.
\end{theorem}

\begin{figure}\label{fig:critical}
  \centering
\includegraphics[scale=0.85]{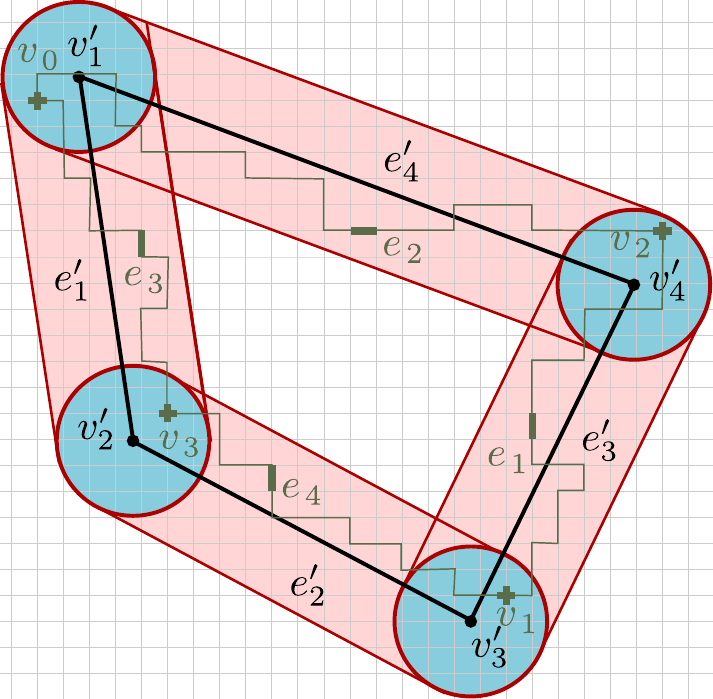}
\caption{A graph $G$ with vertex and edge regions. Critical edges and their
  stable manifolds are shown in green.}
\vspace*{-1ex}
\end{figure}

\begin{hproof}
  We prove the homotopy type by showing that $G$ and $\hat G$ have the same
  first Betti numbers, as the homotopy type of a connected graph is completely
  characterized by its first \\ Betti~number.

  After the termination of \algref{alg:main}, by the assumption on the density
  function, for each graph vertex $v'$ of $G$ we will have exactly one critical
  vertex $v$ of $K$ inside the vertex region of $v'$. This vertex is the local
  maximum of $f$ inside the vertex region of $v'$. For the persistence pairings
  in $P(K)$ with persistence larger than $\delta$, a vertex $v$ of $K$ has to be
  paired with either $+\infty$ or with a critical edge $e$ of $K$ from the edge
  region of a graph edge $e'$ of $G$. And, $e'$ will be incident to $v'$, as
  illustrated in \figref{fig:critical}. Now, for each critical edge $e$ of $K$,
  $e$ must lie inside one of the edge regions of $G$. Moreover, for each each
  $e'$ of $G$ we have exactly one critical edge $e$ of $K$. For the pairings of
  $P(K)$ with persistence larger than $\delta$, each edge $e$ is either paired
  with a vertex $v$ from the vertex region of an incident edge or a triangle $t$
  from the complement of $G^\omega$.

  The one-to-one correspondence of the edges of $G$ and the critical edges of
  $K$ and the vertices of $G$ and the critical vertices of $K$ in $V$, shows
  that the stable manifold of a critical edge $e$ of $K$ that lies in the edge
  region of a graph edge $e'$ of $G$ will be a path in $G^\omega$ joining the
  critical vertices of the vertex regions of the end-points of $e'$. This
  concludes that $\hat G$ and $G$ will have the same first Betti numbers. Also,
  since the critical vertices and edges are localized inside the corresponding
  regions we conclude that $d_H(G,\hat G)<\omega$.
  \sush{other direction.}
\end{hproof}

\section{Discussion}
The nature of our project is ongoing. The noise model discussed in the paper is
only a rough approximation of realistic noise models. We are still in the
process of finding a better noise model. We also hope to find a condition on the
density that enables us to guarantee a small Fr\'echet distance between the
edges of $G$ and the reconstruction.

\paragraph*{Acknowledgments}
The authors acknowledge the generous support of the National Science Foundation
under grants CCF-1618469 and CCF-1618605.  The authors also thank Yusu Wang for
her feedback.

\footnotesize{ \bibliographystyle{acm} \bibliography{paper} }
\end{document}